\documentstyle[preprint,eqsecnum,aps,epsfig]{revtex}

\begin{document}
\draft

\title{Propagators of the Jaynes-Cummings model in open systems}
\author{T.W. Chen, C.K. Law and P.T. Leung}
\address{Department of Physics, The Chinese University of Hong Kong, Shatin, Hong Kong SAR,
PR China}
\date{\today}
\maketitle

\begin{abstract}
We present a propagator formalism to investigate the scattering of
photons by a cavity QED system that consists of a single two-level
atom dressed by a leaky optical cavity field. We establish a
diagrammatic method to construct the propagator analytically. This
allows us to determine the quantum state of the scattered photons for
an arbitrary incident photon packet. As an application, we explicitly
solve the problem of a single-photon packet scattered by
an initially excited atom.

\end{abstract}

\pacs{PACS number(s): 42.50.-p, 03.65.Nk, 03.65.Ca}

\narrowtext

\section{Introduction}

The engineering of novel quantum states of photons is a topic of
interest fundamentally and for applications. As is well known in
atomic and particle physics, exotic states could result from
scattering processes. This suggests light scattering by
suitable quantum systems can be an important tool for state
engineering. A common and yet realistic situation in quantum
optics involves the scattering of photon wave packets, comprising
a few photons, from atoms and other photons situated inside an
optical cavity. The atoms and the cavity field together form a
cavity QED system that scatters incident photon packets (Fig. 1).
The main question is: What is the quantum state of the scattered
photons? As the cornerstone of a series of investigations of this
topic, we establish in the present paper a propagator method to
describe the scattering of photon packets from a cavity where a
two-level atom, either in its ground or excited state, is placed
with the companion of some quasi-mode photons. After the interaction,
the input photons and the quasi-mode photons as well leak out of
the cavity and become the output state of the system. The
propagator method proposed here is capable to extract the details
of the output state and thus form a powerful tool to analyze the
physical content underlying the scattering process.

In addition to its potential applications in the study of novel
quantum states, the model considered here is in fact a generalized
variant of  the Jaynes-Cummings model (JCM) \cite{JC}, which has
been a fundamental model in the realm of photon-atom interaction
ever since its first introduction \cite{JC_review}. Not only does
it provide the simplest description of an atom interacting with
quantized fields in optical cavities, JCM is also an important
tool for controlling quantum states \cite{Law1}. One prominent
generalization of the JCM is the inclusion of the leakage effect
of the cavity, which is inevitable in any real experimental setup.
Indeed recent studies in cavity QED have also emphasized the
important role of dissipation effects in quantum information
devices \cite{Knight}.

In conventional approaches to JCM in a leaky cavity it is
customary to consider the space outside the cavity --- the
environment --- as a Markovian bath, followed by solving the
master equation of the reduced  density matrix of the cavity field
\cite{Puri,Gardiner,Gea,Dae,Won}. Thus, energy inside 
the cavity flows unidirectionally to the environment, 
which plays the role of a sink of energy and the information on the 
exact state of the environment is reckoned unimportant and is lost 
during the evolution.

The model discussed in this paper distinguishes itself from other
dissipative JCM's in that the cavity and its environment play an
equal role during the interaction. These two parts, constituting
the ``universe" in consideration, communicate their information
through partially transmitting walls dividing them. Photons inside
the cavity, described in terms of discrete quasi-modes, interact
with the atom and eventually leak out of the cavity. On the other
hand, incident photons characterized by continuous wave number
enter the cavity,  participate in the interactions with the atom,
and leave the cavity in its final state as well. More importantly,
these two kinds of photons lose their respective identities after
entering (or returning to) the environment and interfere to form
novel photon states. It is for this reason that we devote the
current paper to a proper formalism describing this generalized
JCM model.

To this end, we present here a propagator method that solves
directly the quantum state of the whole amalgamated system,
consisting of the atom, the cavity and the environment. Rather
than discrete cavity modes, the atom now couples with the
continuous field modes of the whole system 
\cite{Loudon,Leung,Scully,Kupi}. 
In this sense, the artificial separation between the cavity 
and its environment is completely eliminated, and the effect of 
dissipation can be treated in a more rigorous and fundamental way. 
Once the propagator of the system is known, the exact details of 
the output photons can be obtained immediately without further ado. 
This knowledge is bound to be important for issues related to 
quantum information and quantum measurement.

The pure state approach has previously been used to study
resonance fluorescence \cite{Mollow} and spontaneous atomic decay
in a vacuum cavity, and exact solutions have been found
\cite{Leung,Lewenstein,Uji,Gar}. The major objective in the
present paper is to formulate a comprehensive theory studying the
dynamical response of a two-level atom under the influence of
photons inside the cavity and from the environment. In a recent
paper by the authors \cite{Law2}, a less general case with the
input photons restricted to the cavity mode (i.e., quasi-mode) was
considered and solved by method of Laplace transform. This
restriction is now removed in this paper. By evaluating
the propagator of the system, whose construction is based on a
diagrammatic expansion method, one can readily handle situations
with an arbitrary initial state. Under the
rotating-wave-approximation, the full propagator is a block
diagonal matrix with a sequence of $2 \times 2$ matrices forming
its diagonal block. Each of these matrices is a propagator of the
system in a subspace categorized by the total excitation number
$N$ (to be defined rigorously in Sect. IV), providing a natural
classification of relevant propagators. A construction method
is established here to express the $N$-excitation
propagators in terms of those of the lower excitations. 
Thus, the full propagator of the system can be obtained in a 
systematic way, paving the way for further investigation into the 
quantum state of the output photons.

This paper is organized as follows: In Sec.~II, we describe the
system under study, its normal modes, the atom-field interactions
and obtain the Hamiltonian. An introduction on the propagator
method and the related diagrammatic expansion will be given in
Sec.~III. The general solution for the propagator is derived in
Sec.~IV. We then apply the propagator for $N=2$ to study the
scattering of a single-photon packet from a two-level atom in 
its excited state in Sec.~V. We draw our conclusion in Sec.~VI.

\section{The Model}

\subsection{The cavity and continuous modes}
We consider here a two-sided Fabry-Perot cavity in one-dimensional space. 
The two transmitting mirrors (hereafter referred to as mirrors R and L) 
are placed at $x=\pm L/2$.
The set of continuous modes of such a Fabry-Perot cavity is known 
in the literature \cite{Loudon,Kupi}.
In the following, we will briefly sketch the main result 
for reference.

The two mirrors enclosing the cavity are
modeled by two thin dielectric slabs of thickness $l$ and
refractive indices $n_{\alpha}$ ($\alpha=$L,R).
In the limit $l \rightarrow 0$ and $n_{\alpha} \rightarrow \infty$ 
such that $n_{\alpha}^2 l \rightarrow \mu_{\alpha}$ is finite,
the corresponding complex amplitude reflection 
$r_\alpha$ and transmission coefficients $t_\alpha$ 
($\alpha={\rm L,R}$) are in turn expressed as
\begin{eqnarray}  \label{transmission}
r_\alpha &=& \frac{i k \mu_\alpha}{2 - i k \mu_\alpha} \, , \\
t_\alpha &=& \frac{2}{2 - i k \mu_\alpha} \, ,
\end{eqnarray}
where $k$ is the wave number.

There are two independent sets of modes for the entire system 
(polarization is ignored in the current study), namely, the
left-propagating modes $u_{\rm L} (k,x)$ and the
right-propagating modes $u_{\rm R}(k,x)$, which are given by
\cite{Loudon,Kupi}:
\begin{equation}
\label{Left_mode}
u_{\rm L} (k,x) = \left\{ 
{\begin{array}{*{20}c} {e^{ikx} + R_{\rm L}(k) e^{ - ikx} } \\
{I_{\rm L}(k) e^{ikx} + J_{\rm L}(k) e^{ - ikx} } \\ 
{T_{\rm L}(k) e^{ikx} } \\ 
\end{array}} \right.
\begin{array}{*{20}c} {} \\ {} \\ {} \\ \end{array}
\begin{array}{*{20}c} {} \\ {} \\ {} \\ \end{array}
\begin{array}{*{20}c} {} \\ {} \\ {} \\ \end{array}
\begin{array}{*{20}c} { - \infty < x < - L/2} \\ 
{- L/2 < x < L/2} \\ {L/2 < x < \infty } \\ 
\end{array}
\end{equation}
\begin{equation}
\label{Right_mode}
u_{\rm R} (k,x) = \left\{ 
{\begin{array}{*{20}c} {T_{\rm R}(k) e^{-ikx}} \\ 
{I_{\rm R}(k)e^{-ikx} + J_{\rm R}(k) e^{ ikx} } \\ 
{e^{-ikx} + R_{\rm R}(k) e^{ikx}} \\ 
\end{array}} \right.
\begin{array}{*{20}c} {} \\ {} \\ {} \\ \end{array}
\begin{array}{*{20}c} {} \\ {} \\ {} \\ \end{array}
\begin{array}{*{20}c} {} \\ {} \\ {} \\ \end{array}
\begin{array}{*{20}c} { - \infty < x < - L/2} \\ 
{- L/2 < x < L/2} \\ {L/2 < x < \infty } \\ 
\end{array}
\end{equation}
where
\begin{eqnarray}
\label{RLk}
R_{\rm L}(k) &=&  \{ r_{\rm L} e^{-ikL} + r_{\rm R} 
e^{ikL + 2 i \arg{t_{\rm L}}}\}/D(k) \, , \\
\label{ILk}
I_{\rm L}(k) &=& t_{\rm L}/D(k) \, , \\
\label{JLk}
J_{\rm L}(k) &=& t_{\rm L} r_{\rm R} e^{ikL}/D(k) \, , \\
\label{TLk}
T_{\rm L}(k) &=& T_{\rm R}(k) = t_{\rm L} t_{\rm R} /D(k) \, ,
\end{eqnarray}
with
\begin{equation}  \label{Dk}
D(k) = 1 - r_{\rm L} r_{\rm R} e^{2 i k L} \, .
\end{equation}
Likewise, $I_{\rm R}(k)$, $J_{\rm R}(k) $ and $R_{\rm R}(k) $ 
can be obtained by interchanging the roles of L and R in the 
above equations. The left-luminating modes are shown in 
Fig.~\ref{L_mode_fig}.
The mode functions $u_{\rm L} (k,x)$ and $u_{\rm R}(k,x)$ 
($k \ge 0$) together form a complete orthonormal set in 
$-\infty \le x \le \infty$, satisfying the orthonormal condition:
\begin{equation}
\int^{\infty}_{-\infty} n(x) u_\alpha(k,x) u^{\ast}_{\beta}
(k',x) dx = 2 \pi \delta_{\alpha \beta} \delta (k-k') \, ,
\end{equation}
where $\alpha$,$\beta=$L,R.

The quasi-mode frequencies, ${\bar k}_m$ ($m=1,2,3,\ldots$),
defined by the zeros of $D(k)$, are given explicitly by
\begin{equation}
{\bar k}_m = \frac{i}{2L} {\rm ln}({r_{\rm L}r_{\rm R}}) +
\frac{m \pi}{L} \, .
\end{equation}
For real-valued $k$ close to a quasi-mode frequency, i.e., $|k -
{\bar k}_m| \ll \pi/L$, we have
\begin{equation}
D(k) = -2iL {\large (} \Delta k + i \kappa_{\rm c} {\large)} \, ,
\end{equation}
where $\Delta k = k - k_{\rm c}$, with
\begin{eqnarray}
k_{\rm c} &=& \frac{m \pi}{L} - \frac{1}{2L} 
{\rm Im}[\ln({r_{\rm L} r_{\rm R}})] \, , \\
\kappa_{\rm c} &=& - \frac{1}{2L} 
{\rm Re}[\ln({r_{\rm L}r_{\rm R}})] \, .
\end{eqnarray}
Here $\kappa_{\rm c}$ is the decay rate of the cavity.

\subsection{The two-level atom and interaction Hamiltonian}

\label{TLA coupling with EM-field} 
Consider a system of a two-level atom placed inside the
cavity described above (or other leaky cavities in general) at $x=x_0$. 
The ground-state energy of the atom is arbitrarily taken as zero, 
while the excited-state energy is $\omega _A$ in units of 
$\hbar =c=1$. The full Hamiltonian of the system 
in the rotating-wave-approximation is given by
\begin{eqnarray}  \label{Hamiltonian_1}
H &=&\omega _A\sigma_{+}\sigma_{-}+\int_0^\infty 
k(a_{k{\rm L}}^{\dagger}
a_{k{\rm L}}+a_{k{\rm R}}^{\dagger}a_{k{\rm R}}) dk  \nonumber \\
&&+\int_0^\infty \left\{\left[ g_{{\rm L}}(k)a_{k{\rm L}}+g_{{\rm
R}}(k)a_{k{\rm R}}\right]\sigma _{+}+{\rm h.c.}\right\}  dk \,,
\end{eqnarray}
where $a_{k\alpha }$ and $a_{k\alpha }^{\dag }$ 
($\alpha ={\rm L},{\rm R}$) are the annihilation and 
creation operators of the $k\alpha $-mode photon, $\sigma _{\pm }$ 
are the pseudo-spin flip operators of the atom. 
They satisfy the usual commutation relations:
$\left[a_{k\alpha},a_{k^{\prime}\alpha^{\prime}}^{\dagger}\right] =
\delta(k-k^{\prime})\delta_{\alpha \alpha^{\prime}} \, ,
\left[a_{k\alpha},a_{k^{\prime}\alpha^{\prime}}\right] =
\left[a_{k\alpha}^{\dagger},a_{k^{\prime}
\alpha^{\prime}}^{\dagger}\right]
=\left[ a_{k\alpha},\sigma_{\pm}\right] = \left[
a_{k\alpha}^{\dag},\sigma_{\pm}\right] =0 \,$, and 
$\left\{\sigma_{-} , \sigma_{+} \right\} = 1$. The coupling constant
$g_{\alpha }(k)$ of the atom with the $k\alpha $-mode photon
depends on the dipole moment of the atom and is proportional to
$u_\alpha(k,x_0)$. Here we are particularly interested in the case
where the transition frequency of the atom is close to one of the
resonance frequencies of the cavity, say, $k_{\rm c}$. Hence,
only those continuous modes with frequencies near $k_{\rm c}$
have significant interactions with the atom and in this
single-mode approximation, 
$g_{\alpha }(k) \propto (\Delta k+i\kappa _{\rm c})^{-1}\,$.

Despite that the atom ostensibly couples with both the R and L
modes, it is always possible to use a unitary transformation to
redefine the photon modes so that the atom interacts with only one
set of modes. For concreteness, we consider in the present paper a
symmetric cavity with identical mirrors and the atom being situated 
at its center. However, it can be proved that generalizations to
cases with dissimilar mirrors, arbitrary atomic position and
one-sided cavity are straightforward. For this specific model, we
have $u_{{\rm L}}(k,x_{0})=u_{{\rm R}}(k,x_{0})$, $g_{{\rm
L}}(k)=g_{{\rm R}}(k)$. Accordingly, one can define a new basis
of photons by the unitary transformation
\begin{eqnarray}
\label{ak+}
a_{k +} &=& \frac{1}{\sqrt{2}} {\large (} a_{k {\rm L}} 
+ a_{k {\rm R}} {\large )} \, , \\
\label{ak-}
a_{k -} &=& \frac{1}{\sqrt{2}} {\large (} a_{k {\rm L}} 
- a_{k {\rm R}} {\large )} \, .
\end{eqnarray}
It is readily observed that the ``$-$" modes do not couple to the
atom. In the following discussion, we will ignore the ``$-$"
modes and will focus on the evolution of the ``$+$" modes, with
the ``$+$" index being suppressed. 
Hence, the full Hamiltonian reduces to
\begin{equation}\label{Hamiltonian}
H =\omega _A\sigma _{+}\sigma _{-}+\int_0^\infty k a_{k}^{\dag
}a_{k} dk    +\int_0^\infty {\large [ } g(k)a_{k}\sigma
_{+}+g^{\ast }(k)a_{k}^{\dag }\sigma _{-} {\large ]} dk \,,
\end{equation}
where $a_{k}$ here actually denotes 
$a_{k+}$ and $g(k)=\sqrt {2} g_{\rm R}(k)$.

\section{Propagator and Feynman Diagrams}
\label{feynman_rules}

The dynamical response of a system with a Hamiltonian $H$ is
governed by the retarded Green's function that satisfies the equation:
\begin{equation}
\left( i \frac{d}{dt} - H\right)
K_{+}(t,t^{\prime})=\delta(t-t^{\prime}) \, ,
\end{equation}
and is null for $t < t^{\prime}$. 
Obviously, $K_{+}$ can be given explicitly by 
$K_{+}(t,t^{\prime}) = e^{-iH(t-t^{\prime })}\theta(t-t^{\prime})$,
where $\theta (x)$ is the Heaviside step function. Accordingly,
its  Fourier transform, $G(\omega)$, defined by the relation
\begin{equation}  \label{inverse_fourier}
G(\omega) = -  i\int_{ - \infty }^\infty {K_{+}(t,0)} e^{ i\omega
t} dt \,
\end{equation}
and termed the retarded propagator $G(\omega)$, can be
expressed symbolically as
\begin{equation}
G(\omega) =  \frac{1}{\omega - H } \, ,
\end{equation}
where the prescription 
$\omega = \mathop {\lim}\limits_{\epsilon \rightarrow 0^+} 
(\omega + i \epsilon)$ 
is assumed hereafter. 

To establish a diagrammatic expansion for $G(\omega)$, 
we separate the Hamiltonian in the form:
$H=H_0+V$, where
\begin{equation}
H_0=\omega _A\sigma _{+}\sigma _{-}+\int_0^\infty k a_{k}^{\dag
}a_{k} dk \,
\end{equation}
is the free atom-field Hamiltonian, and
\begin{equation}\label{}
V=\int_0^\infty {\large [ } g(k)a_{k}\sigma _{+}+g^{\ast
}(k)a_{k}^{\dag }\sigma _{-} {\large ]} dk
\end{equation}
represents the interaction between the atom and the field. Hence,
the propagator can formally be expanded in a power series of $V$,
yielding
\begin{eqnarray}  \label{expandH}
\frac{1}{\omega-H_0 - V}&=& \frac{1}{\omega - H_0} 
+ \frac{1}{\omega - H_0} V \frac{1}{\omega - H_0}  \nonumber \\
&& + \frac{1}{\omega - H_0} V \frac{1}{\omega - H_0} 
V \frac{1}{\omega - H_0} + ........ \, .
\end{eqnarray}

The propagator so defined is an operator. In the
energy-eigenstate basis, the transition amplitudes are given by
the matrix elements of the propagator, also referred to as
``propagator" hereafter. The main purpose of this paper is to
calculate these amplitudes by the associated Feynman diagrams as
illustrated in Fig.~\ref{diag_sym}. The basic construction rules
and interpretations of these diagrams are specified as follows:

1. {\it External and Internal Lines}: Photons are represented by
wavy lines labelled by their momenta $k$. Atoms in the excited
and the ground states are respectively represented by solid and
dashed lines .

2. {\it Vertex Factors}: Each vertex contributes a factor $g(k)$ 
(photon absorption) or $g^{\ast}(k)$ (photon emission) to the 
associated amplitude.

3. {\it Free Propagators}: Each segment in between successive
vertices contributes a factor $(\omega-E)^{-1}$ to the associated 
amplitude, where $E$ is the energy of the free Hamiltonian 
(i.e., $H_0$) in this segment.

4.{\it \ Integrate Over Internal Momenta}. 
For each internal momentum $k$, write down a factor $dk$ and 
integrate.

\section{Evaluation of the propagators}
In this section, we evaluate the matrix elements of the
propagator using the diagrammatic rules stated in
Sec.~\ref{feynman_rules}. Owing to the
rotating-wave-approximation, it is readily observed from
Eq.~(\ref{Hamiltonian}) that the total excitation number
\begin{equation}
N=\sigma _{+}\sigma _{-}+\int_0^{\infty} a_{k}^{\dag }a_{k} dk
\end{equation}
is a constant of motion, resulting in vanishing propagators from
an initial state to a final state with different excitation
numbers. Therefore, the full propagator can be represented by an
infinite sequence of $2\times2$ matrices, each characterized by
its excitation number $N$. Hereafter we will, for convenience,
define an energy eigenstate of $H_0$ with an excitation number $N$
by:
\begin{equation}
\label{photon_state}
|p;k_{1},k_{2},\cdots k_{N-p}\rangle = \frac{1}{\sqrt{(N-p)!}}
a_{k_1}^{\dag} a_{k_2}^{\dag} \cdots a_{k_{N-p}}^{\dag}
\sigma_{+}^p |0;\phi\rangle \equiv |p;{\bf K}_{N-p}\rangle \,,
\end{equation}
where $p=0$($1$) if the atom is in its ground (excited) state, and
$|\phi\rangle$ is the vacuum-field state.
The factor $1/\sqrt{(N-p)!}$ is introduced here to take care of the 
multiple-count of the bosonic states in integrations and 
$|p;k_{1},k_{2},\cdots k_{N-p}\rangle$ is not necessarily normalized 
to unity.

In terms of this notation the four propagators with 
excitation number $N$ are
\begin{equation}
G_{pq}^{(N)}(\omega ;{\bf K}_{N-p},{\bf K^{\prime }}_{N-q})\equiv
\left\langle p;{\bf K}_{N-p}\right| \frac{1}{\omega -H}\left| q;
{\bf K^{\prime }}_{N-q}\right\rangle \,,  \label{def_GN00}
\end{equation}
where $p,q=0,1$. As an example, the propagator of zero excitation
number, governing the propagation of a ground-state atom in
vacuum, is trivially given by
\begin{equation}
\label{vacuum_propagator}
G_{00}^{(0)}(\omega ;\phi ,\phi)= \frac{1}{\omega } \, ,
\end{equation}
for there is only one diagram, namely, the free
propagation of the collective vacuum state.

\subsection{Quasi-mode propagators}
\label{quasi_mode_sec} 
Before proceeding to explicit evaluation
of general propagators, we introduce here the concepts of
quasi-mode photon states and quasi-mode propagators.
A normalized quasi-mode single-photon state is defined by
\begin{equation}
|1_{\rm c} \rangle = a_{\rm c}^{\dag} | \phi \rangle \, ,
\end{equation}
where $a_{c}^{\dag}$ is the effective creation operator for the
quasi-mode:
\begin{equation}
\label{def_ac}
a_{\rm c}^{\dag} = \frac{1}{\sqrt{\lambda}} \int_{0}^{+\infty}
dk \, g^{\ast}(k) \, a_{k}^{\dag} \, .
\end{equation}
The quantity $\lambda$ in the normalization constant 
is the coupling strength defined by
\begin{equation}
\lambda = \int_{0}^{+\infty} g(k) g^{\ast}(k) dk \, .
\end{equation}
Physically speaking, the state $| 1_{\rm c} \rangle$ is, in a perturbative sense and also in
the single-mode-approximation, the cavity field set up by the atom
during de-excitation process. Similarly, the atom-field state
with $N$ excitations, where there are $N-p$ quasi-mode photons
and $p$ atomic excitation, is defined by
\begin{equation}
|p;(N-p)_{\rm c}\rangle \equiv \frac{1}{\sqrt{(N-p)!}}(a_{\rm
c}^{\dag })^{(N-p)}\sigma _{+}^{p}|0;\phi\rangle \,.
\end{equation}
We therefore accordingly define the $N$-excitation quasi-mode 
propagator by
\begin{equation}  \label{def_GcN00}
\Phi^{(N)}_{pq}(\omega) = \left\langle p;(N-p)_{\rm c} \right|
\frac{1}{\omega - H} \left| q;(N-q)_{\rm c} \right\rangle \, .
\end{equation}
From the Hamiltonian given in Eq.~(\ref{Hamiltonian}), the
importance of the quasi-mode propagator is readily clear. Aside
from the input and output photons, those present in the
intermediate states are all quasi-mode photons. Thus, these {\em
quasi-mode propagators form the backbone of our theory from which
other propagators can be derived}.

From definition~(\ref{def_GcN00}), it is obvious that the 
vacuum propagator in Eq.~(\ref{vacuum_propagator}) 
is the simplest quasi-mode propagator: 
$\Phi _{00}^{(0)}(\omega ) = 
G_{00}^{(0)}(\omega ;\phi ,\phi)=\omega^{-1}$.
We begin with the quasi-mode propagator of single excitation
\begin{equation}
\Phi_{11}^{(1)}(\omega)=\langle 1;\phi|\frac{1}{\omega -H}
|1;\phi\rangle \, .
\end{equation}
All relevant diagrams are shown in Fig.~\ref{Gee_fig} and the
physical picture indicated by the diagrams is clear. The
excited-state atom may freely propagate in vacuum, possibly
followed by equal numbers of emissions and absorptions of
quasi-mode photons, and exits in its excited state. In
Fig.~\ref{Gee_fig}, bold-wavy lines are used to represent
intermediate quasi-mode photon state, to distinguish it from the
input and output normal-mode states.

According to the Feynman rules, $\Phi_{11}^{(1)}(\omega)$ 
is given by an infinite series:
\begin{equation}
\Phi_{11}^{(1)}(\omega)=\frac{1}{{\omega -\omega _A}}
\left( {1+\frac{\zeta }{\omega -\omega _A}+\frac{\zeta ^{2}}
{(\omega -\omega _A)^{2}}+\frac{\zeta ^{3}}
{(\omega -\omega _A)^{3}}+\cdots }\right)
\label{x_series} \\
\,,
\end{equation}
where
\begin{equation}
\zeta(\omega) =\int_{0}^{+\infty} \frac{g(k)g^{\ast }(k)}{\omega
-k} dk \,.
\end{equation}
Adopting the single-mode approximation and, as usual, extending
the lower limit of all the $k$-integrations from $0$ to
$-\infty$, we find
\begin{equation}
\zeta (\omega)=\frac{\lambda }{\omega -k_{\rm c}+i\kappa _{\rm c}} \,.
\end{equation}
Therefore, Eq.~(\ref{x_series}) can be expressed in the closed form
\begin{equation}
\label{1st_quasi}
\Phi_{11}^{(1)}(\omega)=\frac{A_{+}^{(1)}}{\omega -\Omega
_{+}^{(1)}}+\frac{A_{-}^{(1)}}{\omega -\Omega _{-}^{(1)}}\,,
\label{Ge0e0}
\end{equation}
where
\begin{eqnarray}  \label{Omega_eN}
A_{\pm }^{(N)} &=& \frac{1}{2}\left[ 1\pm \frac{(\omega _A
-k_{\rm c}+i\kappa_{\rm c})/2}{\sqrt{(\omega _A-k_{\rm
c}+i\kappa
_{\rm c})^{2}/4+N\lambda }}\right] \, , \\
\Omega _{\pm }^{(N)} &=& \frac{\omega _A}{2}+\left( N-\frac{1}{2}
\right)\left( k_{\rm c}-i\kappa _{\rm c}\right) \pm 
\sqrt{\left( \frac{\omega_A-k_{\rm c}
+i\kappa _{\rm c}}{2}\right) ^{2}+N\lambda} \, ,
\end{eqnarray}
for $N=1,2,3,\ldots$. It is then obvious that $\sqrt{N \lambda}$ 
essentially plays the role of the $N$-photon Rabi frequency.

We have derived in the Appendix the quasi-mode propagator 
for an arbitrary excitation number $N$. They are given by:
\begin{eqnarray}  \label{phi10=phi01}
{\Phi }_{11}^{(N)}(\omega ) &=& \frac{A_{+}^{(N)}}
{\omega -\Omega _{+}^{(N)}}+\frac{A_{-}^{(N)}}
{\omega -\Omega _{-}^{(N)}} \, , \\
{\Phi}^{(N)}_{00} (\omega) &=& \frac{1-A^{(N)}_{+}}
{\omega - \Omega^{(N)}_{+}} + \frac{1-A^{(N)}_{-}}
{\omega - \Omega^{(N)}_{-}} \, , \\
{\Phi}^{(N)}_{01}(\omega) &=& \frac{\sqrt{N\lambda}}
{\omega-\omega_A -(N-1)(k_{\rm c} - i \kappa_{\rm c})} 
{\Phi}^{(N)}_{00}(\omega) \, , \\
{\Phi }_{10}^{(N)}(\omega ) &=& {\Phi }_{01}^{(N)}(\omega ) \, .
\end{eqnarray}

\subsection{Propagators of single excitation ($N=1$)}
\label{first_order}

With the help of the quasi-mode propagators,
we can derive the simplest propagators for the case $N=1$
to manifest the techniques in calculations aided by the
diagrammatic scheme. We begin with the propagator
$G^{(1)}_{11}(\omega;\phi,\phi)=\langle 1;
\phi| {(\omega - H)}^{-1} | 1; \phi \rangle$,
which coincides with the first-order quasi-mode propagator given in
Eq.~(\ref{1st_quasi}), i.e.,
\begin{equation}
G^{(1)}_{11}(\omega;\phi,\phi) = \Phi^{(1)}_{11}(\omega) \, .
\end{equation}

One can, of course, follow the same route in the deviation of $
G_{11}^{(1)}(\omega ;\phi ,\phi )$ to obtain the other three
propagators of single excitation. However, it is clear that once
a particular $N$-excitation propagator is derived, other members
in the same class can be easily obtained in an alternative way by
relating them to the one already obtained, as shown in
Fig.~\ref{first-order}. For example, from
Fig.~\ref{first-order}~(a), the propagator $G_{01}^{(1)}(\omega
;k,\phi )=\langle 0;k|{(\omega - H)}^{-1}|1;\phi\rangle$ can be
related to $G_{11}^{(1)}(\omega ;\phi ,\phi )$ by:
\begin{equation}
G_{01}^{(1)}(\omega ;k,\phi )=\frac{g^{\ast }(k)}{\omega -k}
G_{11}^{(1)}(\omega ;\phi ,\phi )\,.  \label{G_g1k_e0}
\end{equation}
Likewise, we can show from Fig.~\ref{first-order}~(b) 
that the propagator
\begin{equation}
G_{10}^{(1)}(\omega ;\phi ,k)=\langle 1;\phi|\frac{1}{\omega
-H}|0;k\rangle =\frac{g(k)}{\omega -k}G_{11}^{(1)}(\omega ;\phi
,\phi )\,.  \label{G_e0_g1k}
\end{equation}
Eqs.~(\ref{G_g1k_e0}) and (\ref{G_e0_g1k}) also reveal a useful
relation of the propagators, namely, the propagator
$G_{pq}^{(N)}(\omega ;{\bf K}_{N-p},{\bf K^{\prime }}_{N-q})$ can
be obtained from $G_{qp}^{(N)}(\omega ;{\bf K^{\prime
}}_{N-q},{\bf K}_{N-p})$ by simply replacing each $g(k)$ with
$g^{\ast }(k)$, and vice versa, which is obvious from the
diagrammatic scheme and the Feynman rules.

Finally, for the scattering of a photon from the $k^{\prime }$-th
mode to the $k$-th mode by the ground-state atom, the
corresponding propagator is
$G_{00}^{(1)}(\omega ;k,k^{\prime })=
\langle 0;k|{(\omega - H)}^{-1}|0;k^{\prime }\rangle$.
Similarly, from Fig.~\ref{first-order}~(c), we have
\begin{equation}
G_{00}^{(1)}(\omega ;k,k^{\prime })=\frac{\delta (k-k^{\prime
})}{\omega -k^{\prime}}+\frac{g(k)g^{\ast }(k^{\prime })}
{(\omega-k)(\omega -k^{\prime })}G_{11}^{(1)}(\omega ;\phi ,\phi )\,.
\label{G_g1k'_g1k}
\end{equation}

\subsection{Construction of propagators of general excitation number $N$}
In general, an $N$-excitation propagator can be calculated from
those of lower excitation numbers. We will develop here a
systematic approach to evaluate the propagators with $N \ge 2$.

First of all, in some diagrams there may exist ``spectator
photons" that do not interact with the atom in the whole
evolution. These diagrams are said to be factorizable (or
unlinked), and can be straightforwardly related to propagators of
lower excitation numbers in a way to be stated explicitly in the
following discussion. We therefore focus mainly on the {\em
linked} diagrams that all the input and output photons take part
in the interactions. These absorptions and emissions can take
place in any order, resulting in different physical processes.

For each linked diagram we label it with a set of momenta, ${\bf
S} \equiv \{{\tilde{k}}_{1},{\tilde{k}}_{2}, \cdots,
{\tilde{k}}_{2N-p-q}\}$, comprising the time-ordered momenta of
the photons created or annihilated in the process. Here $p$ and
$q$ are integers defined in Eq.~(\ref{def_GN00}). The elements of ${\bf S}$
are taken from $N-p$ output photons in the set ${\bf K}_{N-p}$ and
$N-q$ input photons in the set ${\bf K^{\prime }}_{N-q}$, with no
repetitions. In other words, a particular ${\bf S}$ represents a
particular sequence of the $N-q$ absorptions of the input photons
and $N-p$ emissions of the output photons. For a given ${\bf S}$,
we introduce a quantity ${\cal L}_{pq}^{(N)}(\omega ;{\bf S})$ to
denote the corresponding contributions to the propagator. The
symbol ``${\cal L}$" is used here to refer to ``linked" diagrams.
We note here that this convention should be accompanied by a final
symmetrization of the propagator with respect to ${\bf K}_{N-p}$
and ${\bf K^{\prime }}_{N-q}$.

There are at most only three kinds of photons present in any
segment of the Feynman diagrams, namely, the input and output
photons in continuous modes, and the quasi-mode photons. Between
the $2N-p-q$ vertices associated with the sequence defined by
${\bf S}$, the atom only interacts with the quasi-mode photons.
Therefore the evolution is governed by the quasi-mode propagators
derived. This explains the importance of the quasi-mode
propagators discussed in Sec.~(\ref{quasi_mode_sec}).

It is found that ${\cal L}_{pq}^{(N)}(\omega;{\bf S})$ can be
written in the following compact form
\begin{equation}  \label{barG=prodX}
{\cal L}_{pq}^{(N)}(\omega ;{\bf S}) =\left( \prod_{i=1}^{N-q}
g({k^{\prime }}_{i})\right) \left( \prod_{i=1}^{N-p}
g^{\ast}(k_{i})\right) \left( \prod_{i=0}^{2N-p-q}
\Phi_{p_{i}q_{i}}^{(N_{i})}(\omega -E_{i})\right) \,.
\end{equation}
The physical picture of the above equation is clear. $N-p$ times
of absorptions of input photons and $N-q$ times of emissions of 
output photons must take place at some time, 
with amplitude given by the first part of the equation.
These events split the whole process into $2N-p-q+1$ segments. In
a particular segment, the atom interacts only with the quasi-mode
photons. This means that the propagator of this segment is
governed by the quasi-mode propagators derived in the last
section. The presence of continuous mode spectator photons in the
$i$-th segment has the sole effect of shifting the frequency
$\omega $ by $E_{i}$, which is the total energy of the
continuous-mode photons present in the $i$-th segment 
\footnote{For example, in Fig.~\ref{Gekek},
consider the first stage of the second term in the summation,
the excited atom interacts with quasi-mode photon
while the input normal-mode photon $k^{\prime}_1$ acts as
spectator photon. The evolution of the atom plus quasi-mode
would be governed by the the single excitation
quasi-mode propagator $\Phi^{(1)}_{11}~(\omega)$
in the absence of $k^{\prime}_1$.
However when $k^{\prime}_1$ is present,
the propagtor of the whole system in this segment becomes
$\Phi^{(1)}_{11}~(~\omega~-~k^{\prime}_1~)$.
}.

While not explicitly shown, the ${\cal L}_{pq}^{(N)}(\omega;{\bf
S})$ in Eq.~(\ref{barG=prodX}) depends on ${\bf S}$ through the
terms $N_{i}$, $p_{i}$, $q_{i}$ and $E_{i}$.

Summing all diagrams corresponding to all possible ${\bf S}$, we
have
\begin{equation}
\label{Lambda}
{\Lambda}_{pq}^{(N)}(\omega ;{\bf K}_{N-p},{\bf K^{\prime }}
_{N-q}) \equiv \sum_{{\bf S}}{\cal L}_{pq}^{(N)}(\omega ;{\bf S})\,,
\end{equation}
which is the propagator that includes all linked diagrams only.

To include the unlinked diagrams, we go back to the case that 
some of the input and output photons act as spectators in the whole 
process. There may be one spectator, which, 
without loss of generality, can be assumed to be 
$k_{N-p}$ (${k^{\prime}}_{N-q}$). The propagation of the
remaining system is governed by ${\Lambda}^{(N-1)}_{pq}$.
Similarly we can have two photons acting as
spectators, which are assumed to be $k_{N-p}$ (${k^{\prime}}_{N-q}$) 
and $k_{N-p-1}$ (${k^{\prime}}_{N-q-1}$), and so on. 
The maximum possible number of spectator photons is 
$M={\rm min}\{N-p,N-q\}$.
Hence we have
\begin{eqnarray}  \label{solution}
G_{pq}^{(N)}(\omega ;{\bf K}_{N-p},{\bf K^{\prime }}_{N-q}) 
&=& \sum_{{\rm sym}}\sum_{j=0}^{M} 
\left[ \left( \prod_{l=0}^{j-1}
\delta (k_{N-p-l}-{k^{\prime}}_{N-q-l})\right) \right.  \nonumber \\
&& \times \left. 
{\Lambda}_{pq}^{(N-j)}(\omega -\sum_{l=0}^{j-1}k_{N-p-l};
{\bf K}_{N-j-p},{\bf K^{\prime }}_{N-j-q}) \right] \,.
\end{eqnarray}
In the above equation, $\mathop \sum \limits_{{\rm sym}}$ denotes the
symmetrization of the expression with respect to the input and
output photons, and ${\Lambda}$ is given by Eq.~(\ref{Lambda}).

\subsection{Example: Propagators with excitation number $N=2$}

The general expression of the propagators is
simple if the excitation number $N$ is not large. In this section
we will derive the four propagators of excitation number $N=2$.

Consider first the propagator $G_{11}^{(2)}(\omega
;k_{1},k_{1}^{\prime }) \equiv \langle 1;k_{1}|( \omega
-H)^{-1}|1;{k^{\prime }}_{1}\rangle$. According to
Eq.~(\ref{solution}), we have
\begin{equation}
G_{11}^{(2)}(\omega ;k_{1},{k^{\prime }}_{1})={\Lambda}
_{11}^{(2)}(\omega ;k_{1},{k^{\prime }}_{1})
+\delta (k_{1}-{k^{\prime }}_{1})
{\Lambda}_{11}^{(1)}(\omega -k_{1};\phi ,\phi )\,,
\end{equation}
where
\begin{equation}
{\Lambda}_{11}^{(2)}(\omega ;k_{1},{k^{\prime }}_{1})=\sum_{{\bf S}}
{\cal L}_{11}^{(2)}(\omega ;{\bf S})\,,  \label{G2bar}
\end{equation}
with ${\bf S}=\{k_{1},{k_{1}}^{\prime }\}$ or
$\{{k_{1}}^{\prime
},k_{1}\}$.
One can readily show that (see Fig.~\ref{Gekek})
\begin{equation}
{\cal L}_{11}^{(2)}(\omega ;{\bf S}=\{k_{1},k_{1}^{\prime }\}) 
=g({k^{\prime }}_{1})g^{\ast }(k_{1})
{\Phi }_{10}^{(1)}(\omega -k_{1}){\Phi }_{11}^{(2)}(\omega )
{\Phi }_{01}^{(1)}(\omega -{k^{\prime}}_{1}) \,
\end{equation}
and
\begin{equation}
{\cal L}_{11}^{(2)}(\omega ;{\bf S} = \{k_{1}^{\prime },k_{1}\})
= g({k^{\prime}}_{1})g^{\ast }(k_{1})
{\Phi }_{11}^{(1)}(\omega -k_{1})
{\ \Phi}_{00}^{(0)}(\omega -k_{1}-{k^{\prime }}_{1})
{\Phi}_{11}^{(1)}(\omega- {k^{\prime }}_{1}) \, .
\end{equation}
The total propagator is hence
\begin{eqnarray}
\nonumber
G_{11}^{(2)}(\omega ;k_{1},{k^{\prime }}_{1}) &=& 
\Phi_{11}^{(1)}(\omega-k_{1})\left\{ \frac{{}}{{}}
\delta (k_{1}- {k_{1}}^{\prime })
+g({k^{\prime }}_{1})g^{\ast }(k_{1})
\Phi_{11}^{(1)}(\omega -{\ k^{\prime}}_{1})\right. \\
\nonumber
&& \times \left. \left[ \frac{1}{\omega -k_{1}-{k_{1}}^{\prime
}}+\frac{\lambda }{ (\omega -k_{1}-k_{\rm c}+i\kappa _{\rm
c})(\omega -{k_{1}}^{\prime }-k_{\rm c}+i\kappa
_{\rm c})}{\Phi }_{11}^{(2)}(\omega )\right] \right\} \, . \\
\label{G211}
\end{eqnarray}

The other three propagators of the same excitation number can be
evaluated similarly. However, as mentioned in Sec.~\ref{first_order}, 
these propagators can be obtained immediately from their relations 
with $G_{11}^{(2)}$. For example, the propagator
$G_{01}^{(2)}(\omega ;k_{1}k_{2},{k^{\prime }}_{1})=
\langle 0;k_{1},k_{2}|{(\omega - H)}^{-1}|1;{k^{\prime }}_{1}\rangle$ 
is given by (see Fig.~\ref{second-order}~(a))
\begin{equation}  \label{G201_211}
G_{01}^{(2)}(\omega ;k_{1}k_{2},{k^{\prime }}_{1}) =
\frac{1}{\omega -k_{1}-k_{2}} {\large [} g^{\ast
}(k_{1})G_{11}^{(2)}(\omega ;k_{2},{k^{\prime }}_{1})+g^{\ast
}(k_{2})G_{11}^{(2)}(\omega ;k_{1},{k^{\prime }}_{1}) {\large ]} 
\,.
\end{equation}

Likewise, we can obtain the remaining two propagators with 
$N=2$ from Figs.~\ref{second-order}~(b) and (c),
and the results are stated below:
\begin{eqnarray}
\nonumber
G_{10}^{(2)}(\omega ;k_{1},{k^{\prime }}_{1}{k^{\prime}}_{2})
&=&\langle 1;k_{1}|\frac{1}{\omega -H}
|0;{k^{\prime}}_{1},{k^{\prime }}_{2}\rangle\\
&=&\frac{1}{\omega -{k^{\prime }}_{1}-{k^{\prime }}_{2}} {\large [} 
g({k^{\prime }}_{1})G_{11}^{(2)}(\omega ;k_{1},{k^{\prime}}_{2})
+g({k^{\prime }}_{2})G_{11}^{(2)}(\omega ;k_{1},{k^{\prime }}_{1}) 
{\large ]} \, . \\
\nonumber
G_{00}^{(2)}(\omega ;k_{1}k_{2},{k^{\prime}}_{1}{k^{\prime }}_{2})
&=&\langle 0;k_{1},k_{2}|\frac{1}{\omega
-H}|0;{k^{\prime }}_{1},{k^{\prime }} _{2}\rangle \\
\nonumber
&=&\frac{1}{\omega -{k^{\prime }}_{1}-{k^{\prime }}_{2}} {\large [} 
\delta({k^{\prime }}_{1}-k_{1})
\delta ({k^{\prime }}_{2}-k_{2})+\delta ({k^{\prime }}_{1}-k_{2})
\delta ({k^{\prime }}_{2}-k_{1}) {\large ]} \\
\nonumber
&&+\frac{1}{\omega -{k^{\prime }}_{1}-{k^{\prime }}_{2}}{\large [} 
g({k^{\prime }}_{1})G_{01}^{(2)}(\omega ;k_{1}k_{2},{k^{\prime }}_{2})
+g({k^{\prime }}_{2})G_{01}^{(2)}(\omega ;k_{1}k_{2},
{k^{\prime }}_{1}){\large]} \,. \\
\end{eqnarray}

\section{Application: Single-atom Single-photon Scattering}

As an application of our method, we study the scattering of a
single-photon wave packet by an excited two-level atom inside the
cavity. In particular, we investigate how the spectral width of the 
incident photon affects the outcome of stimulated emission process. 
To begin, we consider the incident photon initially prepared 
in the ``+" modes as defined in Eq.~(\ref{ak+}). 
Hence the initial state is given by
\begin{equation}
| \psi(t=0) \rangle = \int dk' C(k')a_{k'} ^{\dag} |1;\phi \rangle
\, ,
\end{equation}
where $C(k')$ is the photon amplitude. At a later time $t$,
the state becomes
\begin{equation}
| \psi(t) \rangle = \int dk B(k,t) a_k ^{\dag} | 1; \phi \rangle +
{1 \over \sqrt{2}} \int\!\!\!\int dk_1 \, dk_2 C(k_1,k_2,t)
a_{k_1} ^{\dag} a_{k_2} ^{\dag}| 0; \phi \rangle \, ,
\end{equation}
where the two-photon amplitude $C(k_1,k_2)=C(k_2,k_1)$ satisfies
the normalization condition,
\begin{equation}
\int \!\!\! \int \left| C(k_1,k_2) \right|^2 dk_1 \, dk_2 = 1 \, .
\end{equation}
For simplicity, we will only consider the resonance case:
$k_{\rm c}=\omega_{A}$.

The long time state $| \psi(t \rightarrow \infty)\rangle$ is
determined by the asymptotic behavior of $C(k_1,k_2,t)$. Utilizing
the propagator obtained in Eq.~(\ref{G201_211}), we have
\begin{equation}
\lim_{t\rightarrow \infty} C(k_1,k_2,t) = \lim_{t \rightarrow
\infty} \frac{i}{2 \pi} \int d\omega e^{-i \omega t} \int
dk^{\prime} G^{(2)}_{10}(\omega;k_1 k_2, k^{\prime}) C(k^{\prime})
\, .
\end{equation}
The explicit form is given by
\begin{eqnarray}
\nonumber C(k_1,k_2,t \rightarrow \infty) &\rightarrow&
\frac{1}{\sqrt{2}} e^{-i(k_1 + k_2)t} \bigg\{ g^{\ast}(k_1)
\Phi^{(1)}_{11}(k_1) C(k_2) +g^{\ast}(k_1)g^{\ast}(k_2) \\
\nonumber && \times \bigg[ \Phi^{(1)}_{11}(k_1)
I_1(k_1,k_2)+\frac{\lambda}{k_1-k_{\rm c}+ i \kappa_{\rm c}}
\Phi^{(1)}_{11}(k_1) \Phi^{(2)}_{11}(k_1+k_2) I_2(k_1,k_2) \bigg]
\bigg\} \\ \label{Ck1k2} && + \{ k_1 \leftrightarrow k_2 \} \, ,
\end{eqnarray}
where
\begin{eqnarray}
\label{I1} I_1(k_1,k_2) &=& \lim_{\delta \rightarrow 0^+}
\int_{-\infty}^{\infty} \frac{g(k') \Phi^{(1)}_{11}(k_1 + k_2
-k')}{k_1 - k'+i \delta} C(k')dk' \, , \\ \label{I2}
I_2(k_1,k_2) &=& \int_{-\infty}^{\infty}
\frac{g(k')}{k_1+k_2-k'-k_{\rm c}+i\kappa_{\rm
c}}\Phi^{(1)}_{11}(k_1+k_2-k')C(k')dk' \, ,
\end{eqnarray}
and $\{ k_1 \leftrightarrow k_2 \}$ denotes the previous
expression with $k_1$ and $k_2$ interchanged. It should be noted
that the contour of the $I_1$-integration should be closed in the
lower half-plane.

The main advantage of our formalism is that it can handle any
state of the incident photon. As an example, we consider that 
$C(k')$ is a lorentzian with a peak frequency
equal to the cavity resonance frequency, i.e., 
\begin{equation}
\label{input_no_gk} C(k')=\sqrt{\frac{\kappa_{\rm in}}{\pi}}
\frac{1}{k'-k_{\rm c}+i\kappa_{\rm in}} \, .
\end{equation}
Here the width $\kappa_{\rm in}$ is the spectral width of the incident
photon. The pole is located in the lower-half-plane in order to
ensure that the atom can only ``feel" the photon for $t \ge 0$.

In Fig.~\ref{joint-count-widths} we show the contour-plot of
$|C(k_1,k_2)|^2$ for $\lambda=0.1\kappa_{\rm c}^2$, 
with various spectral widths of incident photons. 
When $\kappa_{\rm in} \gg \gamma_{\rm sp} $
(Fig.~\ref{joint-count-widths}~(a)), the input photon has a very
short pulse duration compared with the decay time of the atom.
Therefore the incident photon is incapable of having sufficient
interactions with the atom. As a result, the output state is
approximately a direct product state of the input photon and the
spontaneously-decayed photon of the atom, which corresponds to a 
``cross" shape in $|C(k_1,k_2)|^2$. A similar effect can be seen 
if the incident photon has a narrow width such that 
$\kappa_{\rm in} \ll\gamma_{\rm sp}$ (Fig.~\ref{joint-count-widths}~(d)). 
In this case, the input photon has a very long duration and so it
participates in the interaction mainly after the atom has reached 
the ground state, causing no interference with the photon emitted 
from spontaneous decay.

However, interesting features show up when $\kappa_{\rm in}$ is
the same order as $\gamma_{\rm sp}$
(Figs.~\ref{joint-count-widths}~(b) and (c)). We see that the
final two-photon amplitude is drastically different. For example,
in Figs.~\ref{joint-count-widths}~(b), there is an unexpected dip
at the center and peaks at $\Delta k\approx \gamma_{\rm sp}$. In
other words, although the frequencies of the input photon and the
photon emitted in spontaneous decay both peak at $k_c$, it is very
unlikely to have two photons with frequencies around $k_c$ in the
output. Instead, the peak frequency has been shifted to $\Delta
k_1, \Delta k_2 \approx \lambda/\kappa_{\rm c}$.

To understand the interference effects shown in
Figs.~\ref{joint-count-widths}~(b) and (c),  we identify the contributions
of relevant diagrams associated with the propagator.
Fig.~\ref{joint-count-paths} (a) shows the contributions solely
from the unlinked diagram in which the incident photon does not
participate in the interactions at all. Obviously, the corresponding
two-photon amplitude disagrees with the exact one. However, a much
better agreement can be achieved if we include just the lowest
order linked diagram given in Fig.~\ref{joint-count-paths} (b).
The interference between the linked diagram and unlinked diagram
produces a two-photon amplitude that is almost the same as the
exact one (Fig.~\ref{joint-count-paths}(c)).

\section{conclusion}

In this paper, we have developed a diagrammatic scheme to construct
the propagator governing the interaction between an atom and
photons in a leaky cavity. Under the assumption that the frequency
dependence of atom-field coupling is a lorentzian,  we found that
the perturbation series is summable and so exact solution can be
obtained for any excitation number $N$. The propagator provides an
analytical tool to investigate the cavity QED effect for the photon 
scattering problem. In particular, the quantum state of output 
photons in continuous modes can be determined explicitly.

Our results are illustrated by the derivation of propagators with
excitation number $N=2$, which are then applied to study the
scattering of a single-photon wave packet from an excited atom. We
found that the spectral width of the incident photon can
significantly modify the final two-photon amplitudes. This occurs
when $\kappa_{\rm in}$ matches the cavity modified atom decay rate.
%in the bad cavity regime.
Our calculations show that two output
photons are entangled in the sense that their spectrum displays
nontrivial correlation. With the aid of Feynman diagrams, we have
identified the essential processes causing the interference.
However, more detailed investigations are needed for a thorough 
understanding of the rich features in cavity QED assisted photon 
scattering problems.

\acknowledgments Our work is supported in part by the Hong Kong
Research Grants Council (grant No: CUHK4282/00P) and a direct
grant (Project ID: 2060150) from the Chinese University of Hong
Kong.
\appendix

\section{Details of the derivation of quasi-mode propagators}

The main idea of the evaluation of the propagators is to move all the
annihilation and creation operators to the right and left side of
the perturbation series, employing the commutation relations, 
followed by noticing the fact that 
$a_{k}|\phi\rangle = \langle \phi | a_{k}^{\dag}=0$ and
$\sigma _{-}^{2}=\sigma _{+}^{2}=0$.

The commutation relations between $a_{\rm c}$, $a_{\rm c}^{\dag }$
with $H_0$ and $V$ are:
\begin{eqnarray}
\label{ac_dagger_V}
\left[a_{\rm c}^{\dag },V\right] &=&-\sqrt{\lambda }\sigma _{+}\,, \\
\label{ac_V}
\left[ a_{\rm c},V\right] &=&\sqrt{\lambda }\sigma _{-}\,, \\
\label{ac_dagger_H0} \frac{1}{\omega -H_{0}}a_{\rm c}^{\dag }
&=&\int_{-\infty }^{+\infty}dk\,g^{\ast
}(k)\,a_{k}^{\dag }\frac{1}{\omega -k-H_{0}}\,, \\
\label{ac_H0}
a_{\rm c}\frac{1}{\omega -H_{0}} &=&\int_{-\infty }^{+\infty }dk\,
g(k)\frac{1}{\omega -k-H_{0}}a_{k} \, .
\end{eqnarray}

\subsubsection{${\Phi}^{(N)}_{00} (\omega)$}

Consider the quasi-mode propagator
\begin{equation}
\Phi _{00}^{(N)}(\omega )=\langle 0;N_{\rm c}|\frac{1}{\omega
-H}|0;N_{\rm c}\rangle\,.
\end{equation}
The atom is initially at its ground state with $N$ photons in
quasi-mode. The propagator measures the probability that the
system remains in the same state finally.

The interaction-free part
can be evaluated by considering Eq.~(\ref{ac_dagger_H0}), which
gives
\begin{eqnarray}
\nonumber
\langle 0;N_{\rm c}|\frac{1}{\omega -H_{0}}|0;N_{\rm c}\rangle
&=&\frac{1}{N!\lambda ^{N}}{\Huge \langle }0;\phi{\Huge |}
\prod_{i=1}^{N}\int g({k^{\prime }}_{i})\,
d{k^{\prime }}_{i}\,a_{{k^{\prime}}_{i}}^{\dag } 
\int g^{\ast }(k_{i})\,dk_{i}\,a_{k_{i}}
{\Huge |}0;\phi{\Huge \rangle}\frac{1}{\omega 
-\sum_{i=1}^{N}k_{i}} \\
&=&\frac{1}{\omega -N(k_{\rm c}-i\kappa _{\rm c})}\,.
\end{eqnarray}

Paths with an odd number of interactions have zero contributions.
And from the commutation relations, it is readily shown that
\begin{eqnarray}
\nonumber
V\frac{1}{\omega -H_{0}}|0;N_{\rm c}\rangle
&=&\frac{1}{\lambda ^{N/2}\sqrt{N!}}\prod_{j\neq i}\int
dk_{j}\,g^{\ast
}(k_{j})\,a_{k_{j}}^{\dag }\sigma _{+}|0;\phi\rangle 
\sum_{i}\frac{\lambda }
{\omega -k_{\rm c}+i\kappa _{\rm c}-\sum_{j\neq i}k_{j}}\,, \\
\end{eqnarray}
and similarly
\begin{eqnarray}
\left( V\frac{1}{\omega -H_{0}}\right) ^{2}|0;N_{\rm c}\rangle
&=&\frac{1}{\lambda ^{N/2}\sqrt{N!}}\sum_{i}\left( \prod_{j\neq
i}\int dk_{j}\,g^{\ast }(k_{j})a_{k_{j}}^{\dag }\right) \int
dk\,g^{\ast }(k)a_{k}^{\dag
}|0;\phi\rangle  \nonumber \\
&&\times \frac{\lambda }
{(\omega -\sum_{j\neq i}k_{j}-k_{\rm c}+i\kappa_{\rm c})
(\omega-\sum_{j\neq i}k_{j}-\omega _A)}  \nonumber \\
&=&\frac{N\lambda }{\lambda ^{N/2}\sqrt{N!}}\left(
\prod_{j=1}^{N}\int
dk_{j}\,g^{\ast }(k_{j})a_{k_{j}}^{\dag }|0;\phi\rangle \right)  
\nonumber \\
&& \times \frac{1}{(\omega -k_{\rm c}+i\kappa _{\rm
c}-\sum_{i=1}^{N-1}k_{j})(\omega -\omega _{\rm
a}-\sum_{i=1}^{N-1}k_{j})}\,.
\end{eqnarray}

Hence
\begin{eqnarray}
\nonumber
&& \langle 0;N_{\rm c}| \frac{1}{\omega-H}V\frac{1}{\omega-H}V
\frac{1}{\omega-H}|0;N_{\rm c}\rangle \\
&=& \frac{N \lambda}{\left[\omega-N (k_{\rm c} - i
\kappa_{\rm c})\right]^2\left[\omega - \omega_A-(N-1)(k_{\rm c} - i
\kappa_{\rm c})\right]} \, .
\end{eqnarray}

It can be shown that the contributions of different paths with
even times of interactions,
\begin{eqnarray}
\nonumber
&& \langle 0; N_{\rm c}| \frac{1}{\omega-H_0} \left(
\frac{V}{\omega-H_0}\right)^{i} | 0;N_{\rm c} \rangle \\
&=& \frac{1}{\left[\omega - N (k_{\rm c} - i \kappa_{\rm c})\right]^{i/2+1}} 
\left[\frac{N \lambda}
{\omega - \omega_A - 
(N-1)(k_{\rm c} - i \kappa_{\rm c})}\right]^{i/2} \, . 
\end{eqnarray}
Hence
\begin{eqnarray}
{\Phi}^{(N)}_{00}(\omega) &=& \frac{1}{\omega - N (k_{\rm c} -i
\kappa_{\rm c})} \left(
1+\zeta+\zeta^2 + \cdots \right)  \nonumber \\
&=& \frac{1}{\left[ \omega - N (k_{\rm c} -i \kappa_{\rm c})\right](1
-\zeta)} \, ,
\end{eqnarray}
where
\begin{equation}
\zeta=\frac{N\lambda}{\left[\omega - N (k_{\rm c} - i \kappa_{\rm
c})\right] \left[ \omega - \omega_A - (N-1) (k_{\rm c} - i \kappa_{\rm
c})\right] } \, .
\end{equation}

Similar to Eq.~(\ref{Ge0e0}), the propagator can be written as a
sum of two lorentzians,
\begin{eqnarray}  \label{GgNgN_2}
{\Phi }_{00}^{(N)}(\omega ) &=&\frac{\omega -\omega _{\rm
a}-(N-1)(k_{\rm c}-i\kappa _{\rm c})}{\left[\omega -N(k_{\rm
c}-i\kappa _{\rm c})\right] \left[\omega -\omega
_A-(N-1)(k_{\rm c}-i\kappa _{\rm c})\right]-N\lambda }  \nonumber \\
&=&\frac{1-A_{+}^{(N)}}{\omega -\Omega
_{+}^{(N)}}+\frac{1-A_{-}^{(N)}}{ \omega -\Omega _{-}^{(N)}}\,,
\end{eqnarray}
where
\begin{eqnarray}
A_{\pm }^{(N)} &=&\frac{1}{2}\left[ 1\pm \frac{(\omega _{\rm
a}-k_{\rm c}+i\kappa _{\rm c})/2}{\sqrt{(\omega _A-k_{\rm
c}+i\kappa _{\rm c})^{2}/4+N\lambda }}\right] \, ,
\label{Omega_gN} \\
\Omega _{\pm }^{(N)} &=&\frac{\omega _A}{2}+\left(
N-\frac{1}{2}\right) \left( k_{\rm c}-i\kappa _{\rm c}\right) \pm
\sqrt{\left( \frac{\omega
_A-k_{\rm c}+i\kappa _{\rm c}}{2}\right) ^{2}+N\lambda }\,.  
\nonumber \\
&&
\end{eqnarray}

By substituting $N=0$, we have $A^{(0)}_{+} = 0$, $A^{(0)}_{-} = 1$, 
$\Omega^{(0)}_{+} = \omega_A - k_{\rm c} + i \kappa_{\rm c}$,
$\Omega^{(0)}_{-} = 0$, and the propagator reduces to
\begin{equation}
{\Phi}^{(0)}_{00} (\omega)= \frac{1}{\omega} \, .
\end{equation}

\subsubsection{${\Phi}^{(N)}_{01} (\omega)$}

The propagator
\begin{equation}
{\Phi}^{(N)}_{01} (\omega) = \langle 0;N_{\rm c} |
\frac{1}{\omega - H} | 1;(N-1)_{\rm c}\rangle
\end{equation}
can be evaluated by the usual commutation relations, and
noticing that
\begin{eqnarray}
\nonumber
&& V\frac{1}{\omega - H_0}\frac{1}{\sqrt{(N-1)!}}
(a_{\rm c}^{\dag})^{N-1}\sigma_+ |0;\phi \rangle \\
&=& \frac{1}{\sqrt{N!}\lambda^{N/2}} 
\left(\prod_{i}^{N}\int dk_i
g^{\ast}(k_i) a_{k_i}^{\dag}| 0;\phi \rangle \right) 
\frac{\sqrt{N\lambda}}{\omega - \omega_A - \sum_{j=1}^{N-1} k_j} 
\, .
\end{eqnarray}
It can be proved that
\begin{eqnarray}
{\Phi}^{(N)}_{01}(\omega) &=& 
\frac{\sqrt{N\lambda}}
{\omega -\omega_A -(N-1) (k_{\rm c} - i \kappa_{\rm c})} 
{\Phi}^N_{00}(\omega)  \nonumber \\
&=& \frac{\sqrt{N\lambda}}{\Omega^{(N0)}_{+1}-\Omega^{(N0)}_{-1}}
\left[ \frac{1}{\omega - \Omega^{(N0)}_{+1}} -\frac{1}{\omega -
\Omega^{(N0)}_{-1}} \right] \, .
\end{eqnarray}

\subsubsection{${\Phi}^{(N)}_{10}(\omega)$}

The calculation of
\begin{equation}
{\Phi}^{(N)}_{10}(\omega) = \langle 1;(N-1)_{\rm c}|
\frac{1}{\omega - H} | 0;N_{\rm c} \rangle
\end{equation}
is similar to that of $\Phi^{(N)}_{01} (\omega)$. In fact, the
symmetries in the commutation relations
result in the relation
\begin{equation}  \label{Geg_Gge}
{\Phi}^{(N)}_{10}(\omega)={\Phi}^{(N)}_{01} (\omega) \, .
\end{equation}

The above equation can also be obtained immediately by replacing
$g(k)$ with $g^{\ast}(k)$, which does not change the expression
in this case since the quasi-mode propagators do not contain any
free $g(k)$ or $g^{\ast}(k)$ terms.

\subsubsection{${\Phi}^{(N)}_{11} (\omega)$}

The propagator
\begin{equation}
{\Phi}^{(N)}_{11} (\omega) = \langle 1;(N-1)_{\rm c}|
\frac{1}{\omega - H} | 1;(N-1)_{\rm c}\rangle
\end{equation}
can be evaluated by similar derivations, yielding
\begin{equation}
\langle 1;(N-1)_{\rm c} | \frac{1}{\omega - H_0} | 1;(N-1)_{\rm c} 
\rangle = \frac{1}{\omega - \omega_A - (N-1) 
(k_{\rm c} -i \kappa_{\rm c})} \, ,
\end{equation}
and for an even integer $i$,
\begin{eqnarray}
\nonumber
&& \langle 1; (N-1)_{\rm c}| \frac{1}{\omega-H_0}
\left(\frac{V}{\omega -H_0} \right)^{i} | 1; (N-1)_{\rm c} \rangle \\
&=& \frac{N\lambda}
{\left[\omega - \omega_A - (N-1) (k_{\rm c} - i \kappa_{\rm c})\right]^2}
\langle 0;N_{\rm c}| \frac{1}{\omega - H_0}
\left( \frac{V}{\omega - H_0} \right)^{i-2} |0;N_{\rm c} \rangle \, .
\end{eqnarray}
Similarly all paths with odd number times of interactions have
null contributions to the propagator. Hence, we have
\begin{eqnarray}
\nonumber
{\Phi}^{(N)}_{11} (\omega)  \nonumber &=& 
\frac{1}{\omega - \omega_A - (N-1) (k_{\rm c} - i \kappa_{\rm c})} \\
\nonumber
&& \times \left\{ 1 + \frac{N\lambda}
{\left[\omega - N (k_{\rm c} - i\kappa_{\rm c})\right]
\left[\omega -\omega_a - (N-1) 
(k_{\rm c} - i \kappa_{\rm c})\right] - N\lambda}\right\} \\
\nonumber
&=& \frac{\omega - N(k_{\rm c} - i \kappa_{\rm c})}
{(\omega-\Omega^{(N)}_{+})
(\omega - \Omega^{(N)}_{-})} \\
\label{GeNeN_2}
&=& \frac{A^{(N)}_{+}}{\omega - \Omega^{(N)}_{+}} + 
\frac{A^{(N)}_{-}}{\omega - \Omega^{(N)}_{-}} \, .
\end{eqnarray}

A very simple relation exists between the
$e$(excited-state)$\rightarrow e$ and
$g$(ground-state)$\rightarrow g$ propagators,
\begin{equation}  \label{Ggg_Gee}
{\Phi}^{(N)}_{11} (\omega) = \frac{\omega - N (k_{\rm c} - i
\kappa_{\rm c})}{\omega - \omega_A - (N-1) (k_{\rm c} - i
\kappa_{\rm c})} {\Phi}^{(N)}_{00}(\omega) \, .
\end{equation}
It can be verified that ${\Phi}^{(N)}_{11}(\omega)$ 
reduces to that of Eq.~(\ref{Ge0e0}) if we take $N=1$.

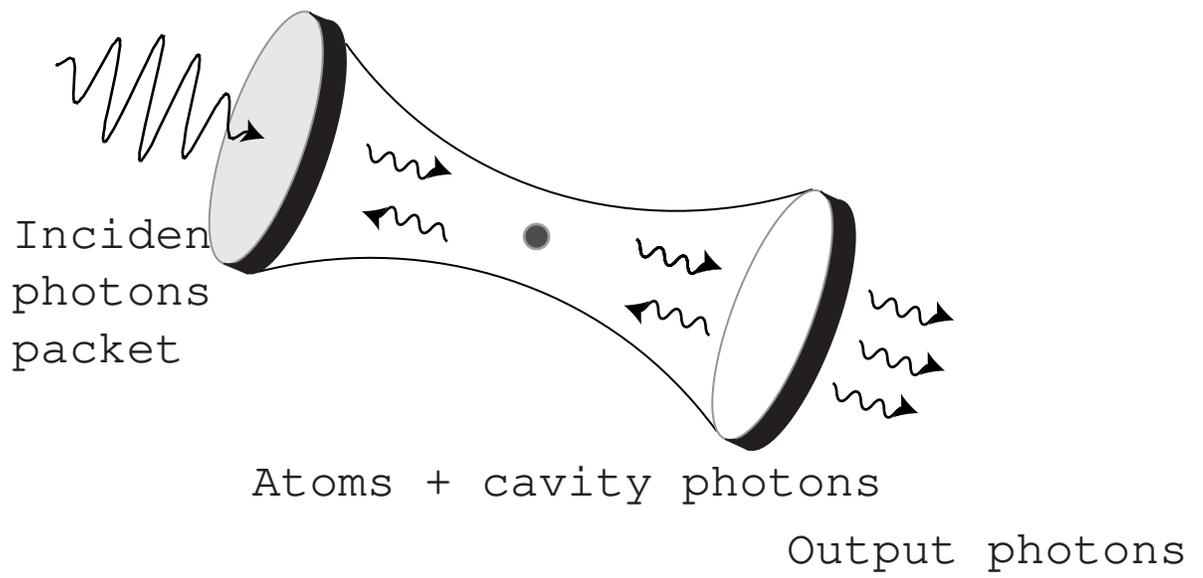
\begin{figure}[tbp]
\caption{A sketch of the system: A two-sided Fabry-Perot cavity with
a two-level atom inside and partially reflecting mirrors at both
ends. We only consider identical mirrors and the atom being locating at
the center.} \label{sketch}
\end{figure}

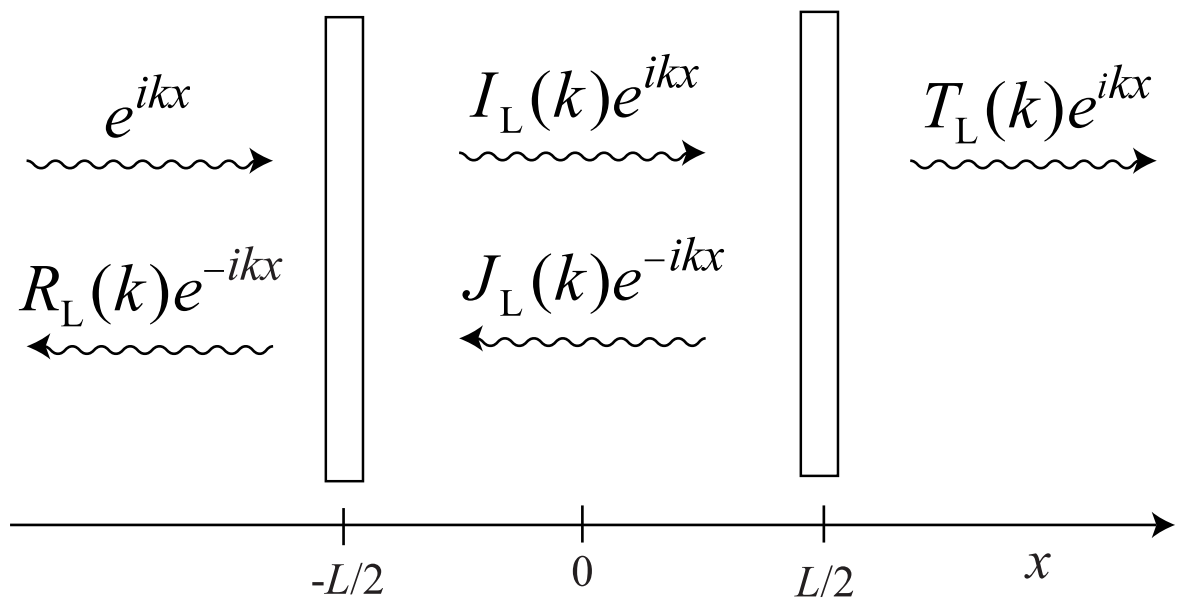
\begin{figure}[tbp]
\caption{The left-luminating modes, with lumination from the left, 
and transmitted waves only at the right. This set of modes
are labeled by the subscript L and the positve wave-number $k$. 
Together with the right-luminating modes, these continous
field modes form a complete and orthogonal set of the system.} 
\label{L_mode_fig}
\end{figure}

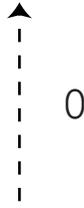
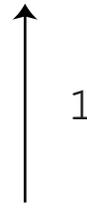
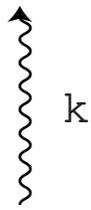
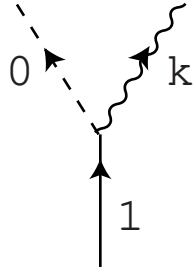
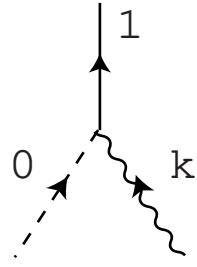
\begin{figure}[tbp]
\caption{Basic components of the Feynman diagrams: 
(a) The total propagator from initial state $\psi_i$ to final state 
$\psi_f$. 
(b) Free propagation of a ground-state atom. 
(c) Free propagation of an excited-state atom.
(d) Free propagation of a $k$-th mode photon. 
(e) An excited-state atom decays into
ground state and emits a $k$-th mode photon. 
(f) A ground-state atom excited by a $k$-th mode photon 
and jumps to the excited state.} \label{diag_sym}
\end{figure}

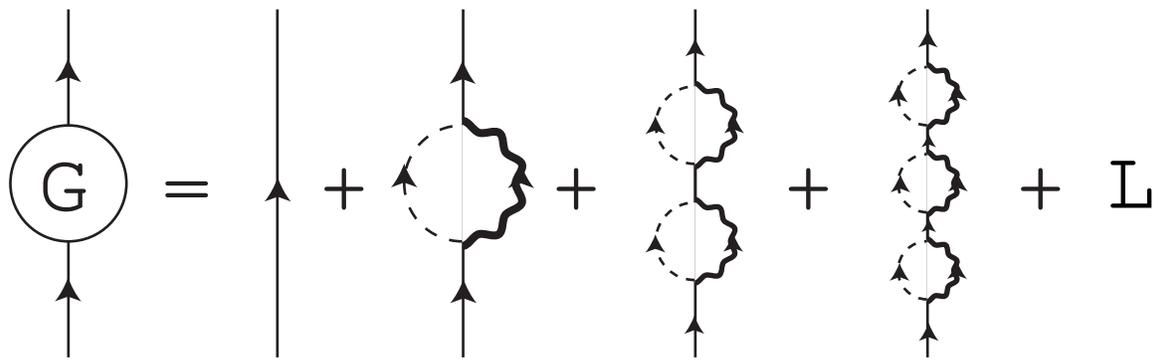
\begin{figure}[tbp]
\caption{The propagator $\Phi^{(1)}_{11}(\omega)
=G^{(1)}_{11}(\omega;\phi,\phi)$ can be expressed as the sum of 
an infinite series, corresponding to all the possible paths.
Bold-wavy lines are used to represent
intermediate quasi-mode photon state. 
This series is exactly summable.
} \label{Gee_fig}
\end{figure}

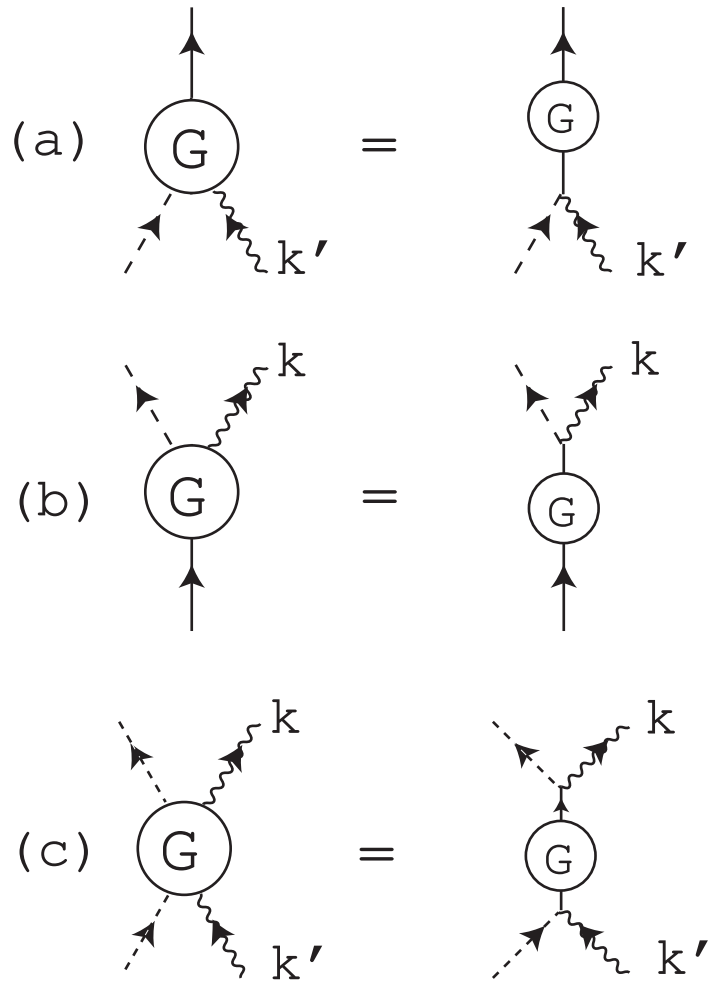
\begin{figure}[tbp]
\caption{Relationship between the propagators of
single-excitation. The $G$'s appear in different components are 
different because they are attached with different input 
and output legs.} 
\label{first-order}
\end{figure}

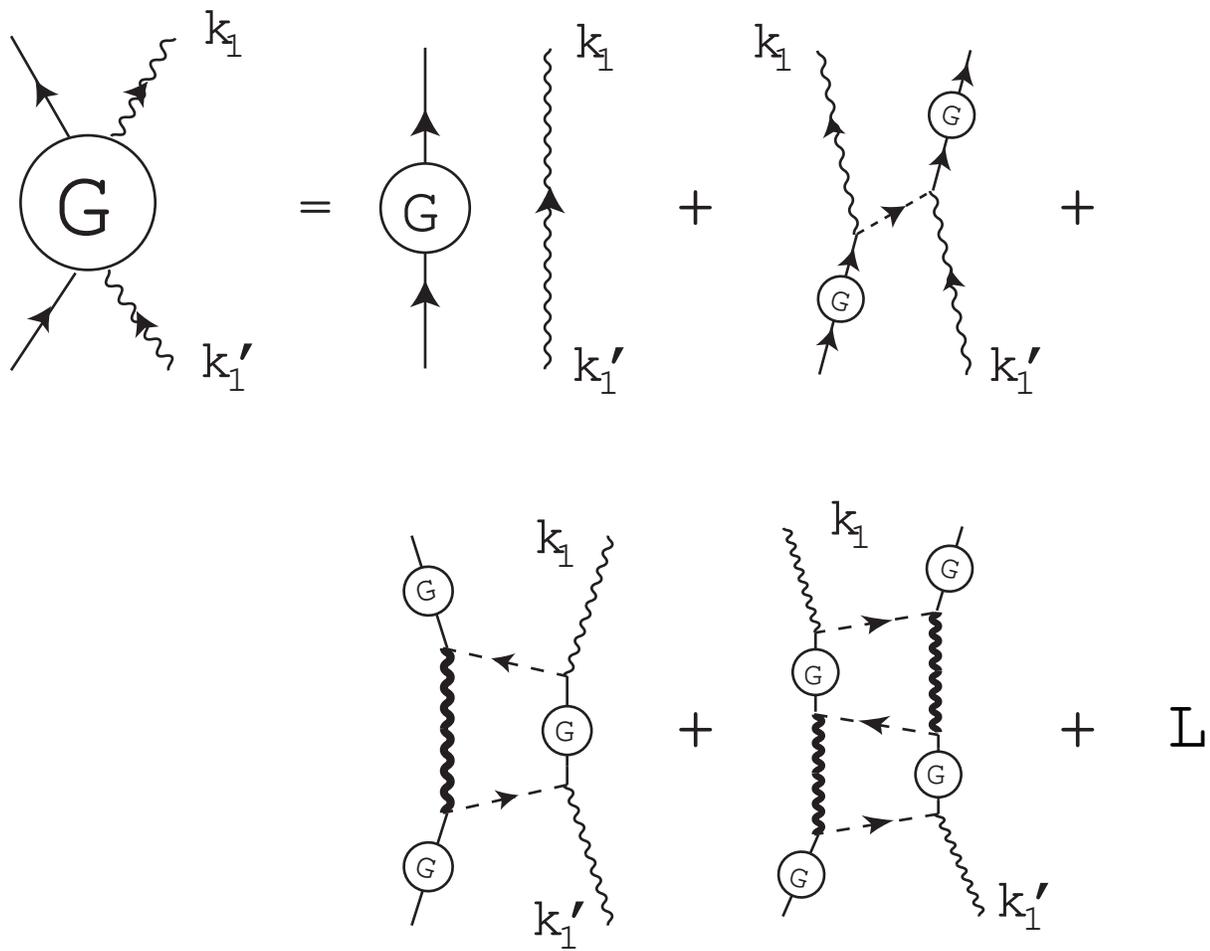
\begin{figure}[tbp]
\caption{Feynman diagrams for 
$G^{(2)}_{11}(\omega;k_1,{k^{\prime}}_1)$.
The first diagram is the unlinked diagram with a spectator photon.
The second diagram corresponds to 
${\bf {\rm S}}=\{ k_1,k^{\prime}_1\}$, 
which means the atom emits the output photon prior to the 
absorption of the input. All other diagrams belong to the 
third group, corresponding to
${\bf {\rm S}}=\{ k^{\prime}_1, k_1\}$.} \label{Gekek}
\end{figure}

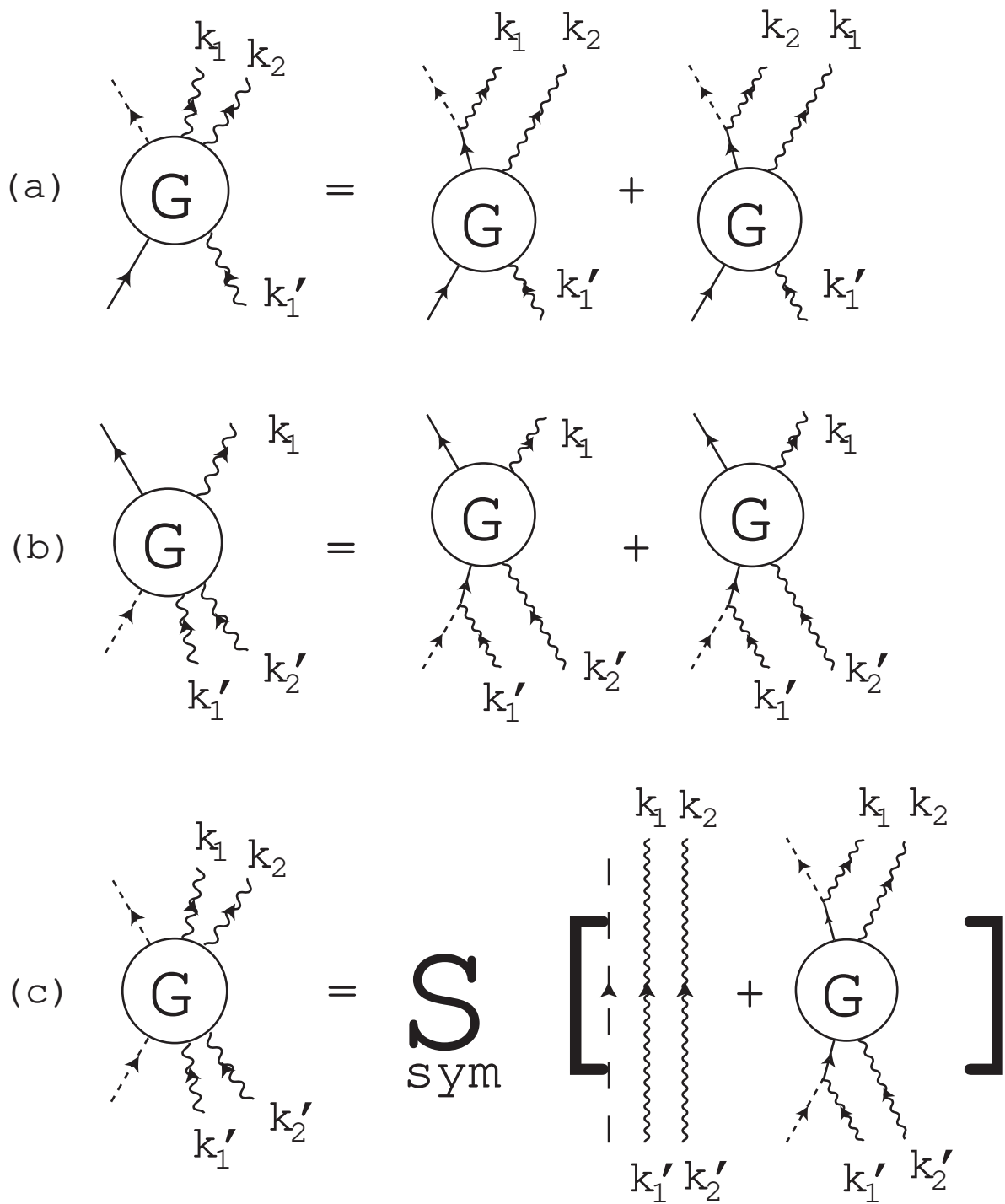
\begin{figure}[tbp]
\caption{Relationship between the propagators with two excitations
$N=2$.} \label{second-order}
\end{figure}

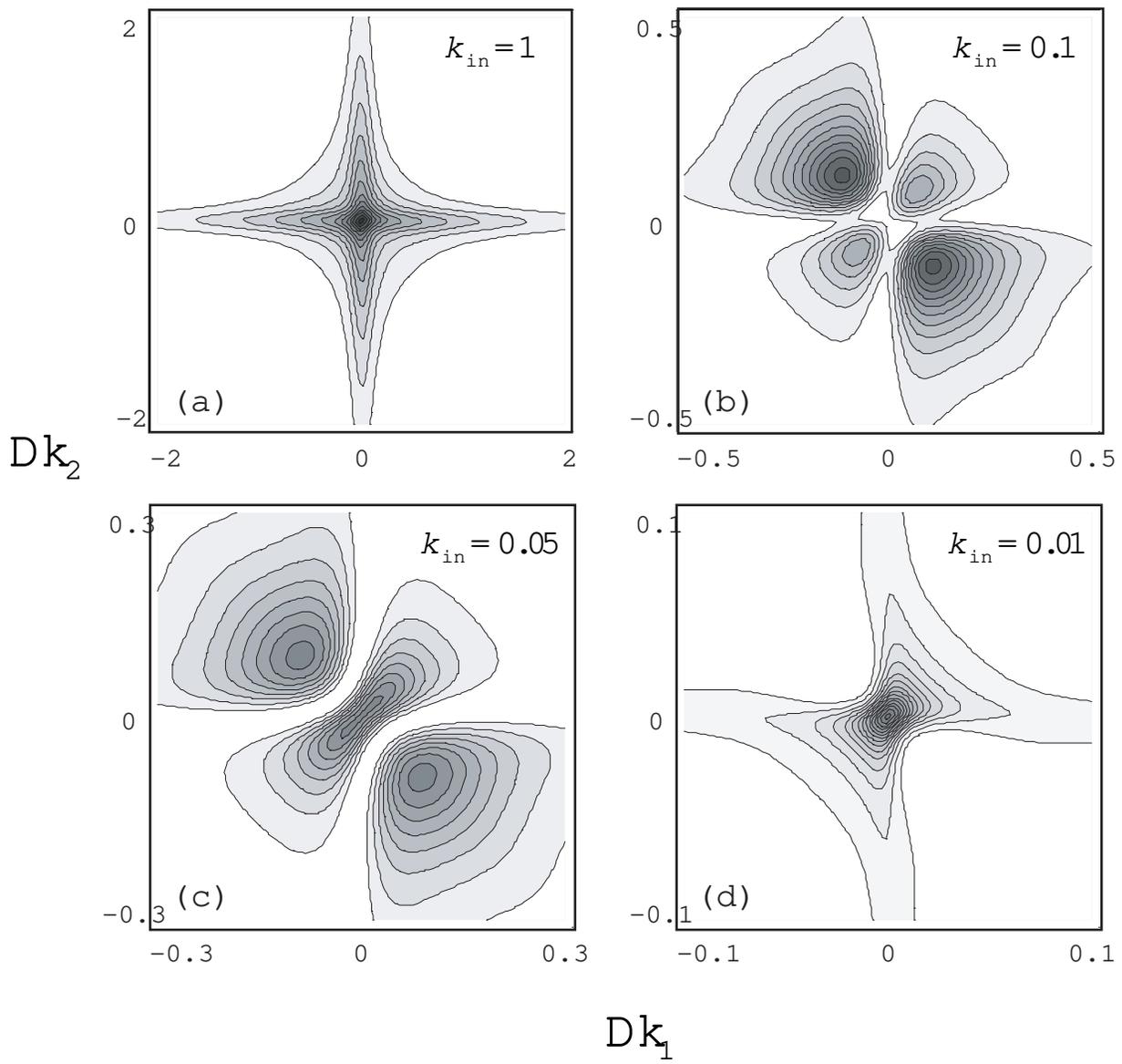
\begin{figure}[tbp]
\caption{Contour-plot of $\left| C(k_1,k_2) \right|^2$ for $\lambda=0.1 
\kappa_{\rm c}^2$, with four different widths of the input photon: 
(a) $\kappa_{\rm in}= 10 \gamma_{\rm sp}$, 
(b) $\kappa_{\rm in}=\gamma_{\rm sp}$, 
(c) $\kappa_{\rm in}= 0.5 \gamma_{\rm sp}$ and 
(d) $\kappa_{\rm in}=0.1 \gamma_{\rm sp}$. 
The axis labels are in unit of $\kappa_{\rm c}$.}
\label{joint-count-widths}
\end{figure}

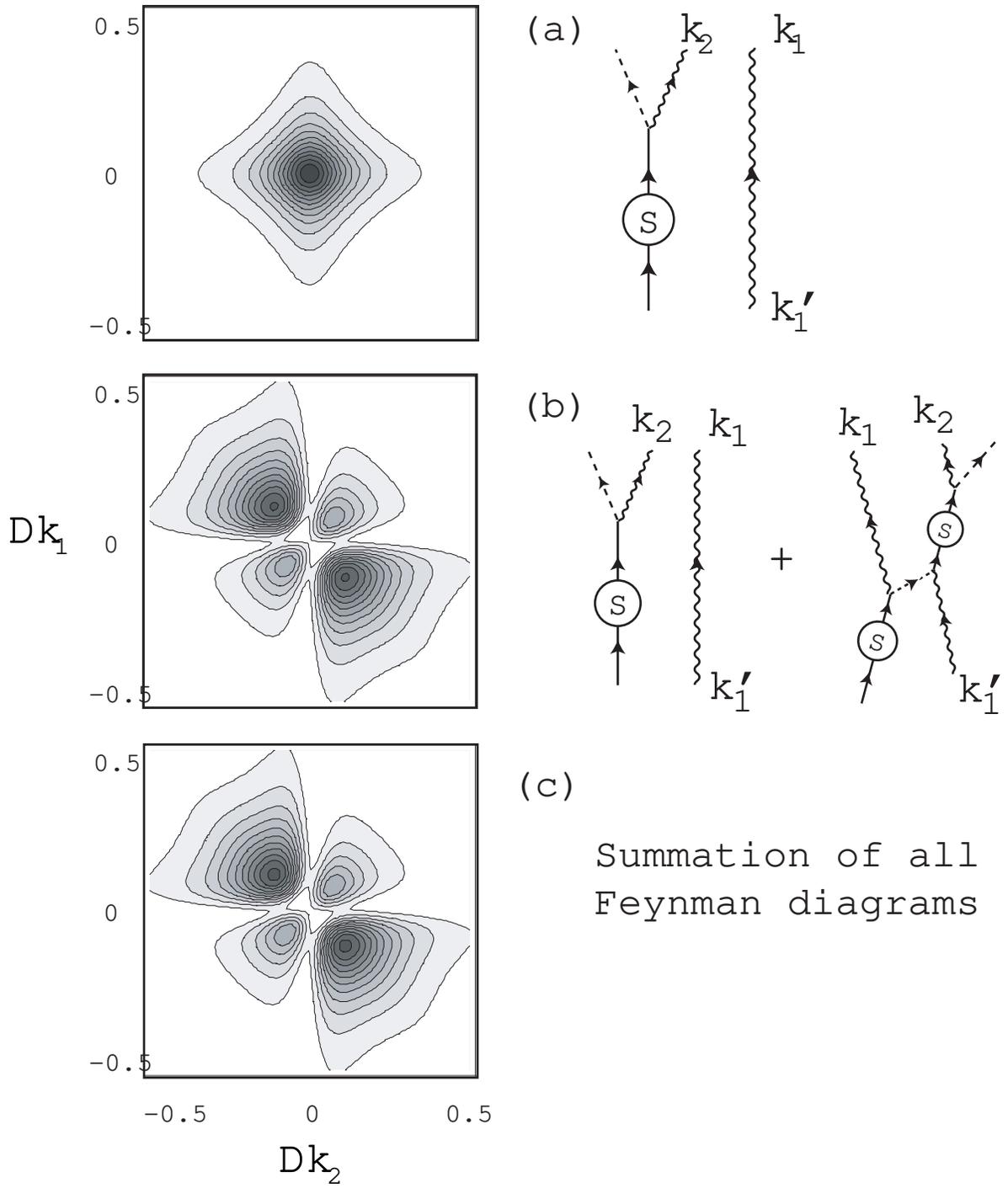
\begin{figure}[tbp]
\caption{The contributions of different Feynman diagrams for the 
case $\kappa_{\rm in} = \gamma_{\rm sp}$
in Fig.~\ref{joint-count-widths}~(b):
(a) The unlinked diagram.
(b) The lowest order linked diagram plus the unlinked diagram.
(c) All diagrams.
The axis labels are in unit of $\kappa_{\rm c}$.} 
\label{joint-count-paths}
\end{figure}

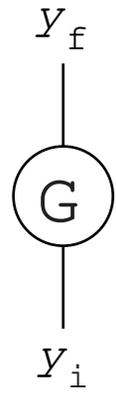
\end{document}